\documentclass[12pt]{article}
\usepackage[english]{babel}
 \usepackage[cp1251]{inputenc}
 \usepackage{eufrak}
\begin{document}
 \sloppy

\title{Squares of White Noise, \( SL(2,\mathbf{C})\) and \newline Kubo
-- Martin -- Schwinger States }

\author{D.V. Prokhorenko \footnote{Institute of Spectroscopy, RAS 142190 Moskow Region, Troitsk}} \maketitle
\begin{abstract}
We investigate the structure of Kubo
--- Martin --- Schwinger (KMS) states on some extension of the universal
enveloping algebra of \(SL(2,\mathbf{C})\). We find that there
exists a one-to-one correspondence between the set of all
covariant KMS states on this algebra and the set of all
probability measures \(d\mu\) on the real half-line
\([0,+\infty)\), which decrease faster than any inverse
polynomial. This problem is connected to the problem of KMS states
on square of white noise algebra.
\end{abstract}
\newpage
\section{Introduction.}
The basis object in quantum field theory is \(S\)-matrix [1,2],
which describes the scattering problem at infinite times. But in a
number of cases one is interested in the behaviour of quantum
systems at large, but finite intervals of time. The general method
for studying dynamical problem in quantum field theory is the
method of stochastic limit developed by L. Accardy, I.V. Volovich
and others [3]. This method leads to quantum stochastic equation.
\newline
\indent In [4] the authors have studied quantum stochastic
equations of the form\newline
\begin{equation}
i\frac{d}{d\tau}U_\tau=(a((b^+_\tau)^2+b^2_\tau)+cb^+_\tau
b_\tau)U_\tau,
\end{equation}
where \(a,b\) are real numbers and \(\{b^+_\tau, b_\tau\}\) ---
quantum white noise, i.e. the pair of operator-valued
distributions satisfying the canonical commutation relations:
\begin{eqnarray}
{[b_\tau,b_{\tau'}]=[b_\tau^+,b_{\tau'}^+]=0},\nonumber\\
{[b_\tau,b^+_{\tau'}]=\delta(\tau-\tau')}.
\end{eqnarray}
\indent These equations contains the squares of white noise. After
the renormalization suggested in [4] these squares generate
so-called square of white noise (SWN) algebra [5,6]. KMS states
was introduced in [7]. KMS states on SWN algebra were considered
in [8], where it was found the example of KMS state on SWN
algebra. Our main goal is to clarify the structure of KMS states
on SWN algebra. After discretising suggested in section 2 our
problem will reduce to analogous problem for
\(U(\mathfrak{sl}(2,\mathbf{C}))\),  the universal enveloping
algebra of  \(\mathfrak{sl}(2,\mathbf{C})\), where
\(\mathfrak{sl}(2,\mathbf{C})\) is the Lie algebra of
\(SL(2,\mathbf{C})\). In the present paper we give complete
description of KMS states on some extension of
\(\mathfrak{sl}(2,\mathbf{C})\). We find that there exists a
one-to-one correspondence between the set of all KMS states on
this algebra and the set of all probability measures \(d\mu\) on
the real half-line \([0,+\infty)\), which decrease faster than any
inverse polynomial. Our main result contained in Theorem 1.
\section{ Problem setup.}
In this section the necessary notion are introduced and the main
result (Theorem 1) is formulated. Let \(\Gamma\) be a space of
piecewise continuous functions on \([0,1]\). \(\Gamma\) is a
Hilbert algebra with respect to the complex conjugation and the
scalar product of the form
\begin{equation}
\langle f|g\rangle=\int \limits_0^1 f^\star(x)g(x)dx.
\end{equation}
\indent Let \(\mathcal{B}\) be a \(\star\)-algebra generated by
 \begin{eqnarray}
 B(f),B^+(f),N(f),\; f \in \Gamma \nonumber
 \end{eqnarray}
 with the following relations: \newline
a)\(B(f)\)  is an antilinear functional of  \(f\),
  \(B^+(f)\) is a linear functional of \(f\),
  \(N(f)\) is a linear functional of \(f\).\newline
b)
 \begin{eqnarray}
{ [B(f),B^{+}(g)]}=2N(f^{\star}g),\nonumber\\
 {[B(f),N(g)]=2B(f^{\star}g)},\nonumber\\
 {[N(f),N(g)]=0}.
 \end{eqnarray}
 The involution is defined by the formulas
\begin{eqnarray}
(N(f))^\star=N(f^\star),\nonumber\\
(B(f))^\star=B^+(f).
\end{eqnarray}
Algebra \(\mathcal{B}\) is called square of white noise algebra
[5,6].
\newline
 \indent Let \(\mathcal{E}\) be a
\(\star\)-algebra with a unit. The state \(\tau\) on
\(\mathcal{E}\) is a positive linear functional satisfying the
following condition: \(\tau(1)=1\).\newline
 \indent Let \(\mathcal{E}\) be a
 \(\star\)-algebra, \(\beta\) be a real positive number and \(V_t\) (\(t \in \mathbf{R}\))
  be an one-parameter group of its automorphisms. We
say, that the linear functional \(\tau\) on \(\mathcal{E}\) is a
KMS-functional with respect the pair \(\{\beta, V_t\}\) if
\(\forall A,B \in \mathcal{E}\) there exists continuous function
\(F_{AB}:S_\beta\rightarrow\mathbf{C}\) which is holomorphic
inside the strip \(S_\beta=\{z \in\mathbf{C}\mid 0\leq \rm Im \mit
z\leq\beta\}\) such that for any real \(t\)
\begin{eqnarray}
F_{AB}(t)=\tau(AV_{t}(B))
\end{eqnarray}
and
\begin{eqnarray}
\tau(V_t(A)B)=F_{BA}(t+i\beta).
\end{eqnarray}
Let \(U_t\), \(t \in \mathbf{C}\) be an one-parameter group of
automorphisms of \( \mathcal{B}\), defined by the following
relations:
\begin{eqnarray}
U_t(B^+(f))=B^+(fe^{it\omega}),\nonumber\\
U_t(B(f))=B(fe^{it\omega}),\nonumber\\
U_t(N(f))=N(f).
\end{eqnarray}
Here \(\omega(x)\) is a real-valued positive continuous function
on \([0,1]\).\newline
 \indent Our aim is to classify all KMS states
on \(\mathcal{B}\) with respect the pair \(\{\beta, U_t\}\) . The
discrete variant of our problem is to classify all KMS states on
the algebra \(\mathcal{C}^m\), generated by generators
\(B_i,B^+_i, N_i\), satisfying the following relations:
\begin{eqnarray}
{[B_i,B_j]=[B^+_i,B^+_j]=[N_i,N_j]=0},\nonumber\\
{ [B_i,B^+_j]=2\delta_{i,j} N_i},\nonumber\\
 {[B_i,N_j]=2\delta_{i,j}B_i},\nonumber\\
 i,j=1,...,m.
\end{eqnarray}
\indent The KMS condition has the form:
\begin{eqnarray}
\tau(AB)=\tau(BU_{i\beta}(A)),\; A,B \in \mathcal{C}^m,
\end{eqnarray}
here \(U_t\) acts on generators as
\begin{eqnarray}
U_t(B^+_i)=B^+_ie^{it\omega_i},\nonumber\\
U_t(B_i)=B_ie^{-it\omega_i},\nonumber\\
U_t(N_i)=N_i,\nonumber\\
\omega_i>0, \;i,j=1,...,m.
\end{eqnarray}

\indent In the first instance we consider the case of the algebra
\(\mathcal{C}^m\) for \(m=1\). Remind that
\(\mathfrak{sl}(2,\mathbf{C})\), the Lee algebra of
\(SL(2,\mathbf{C})\)  is generated by generators \(X,Y,H\) with
the following relations:
 \begin{eqnarray}
 [X,Y]=H,\nonumber\\
{ [X,H]=-2X},\nonumber\\
 {[Y,H]=2Y}.
 \end{eqnarray}
 Using substitution
\begin{eqnarray}
\frac{1}{\sqrt{2}}B_i=Y_i,\nonumber\\
\frac{1}{\sqrt{2}}B^+_i=-X_i,\nonumber\\
  H_i=N_i
\end{eqnarray}
we see that the algebra \(\mathcal{C}^1\) coincide with the
universal enveloping algebra of \(\mathfrak{sl}(2,\mathbf{C})\)
with the involution of the form
\begin{eqnarray}
H^\star=H,\nonumber\\
X^\star=-Y.
\end{eqnarray}
\indent Denote by \(\mathcal{P}\) the set of all continuous
complex-valued function on \(\mathbf{R}\) which increase slowly
than some polynomial at infinity. Let \(a\) be a real number.
Denote by \(T_a\) an operator acting in the space \(\mathcal{P}\)
as follows\newline
\begin{eqnarray}
T_a: f(x)\mapsto(T_a f)(x)=f(x-a).
\end{eqnarray}
\indent \textbf{Definition.} Denote by \(\mathcal{A}\)  a
\(\star\)-algebra, generated by generators \(X, Y, N_F, \; F \in
\mathcal{P}\) which satisfy  the following relations:
\begin{eqnarray}
N_{\lambda F+\mu G}=\lambda N_F+\mu N_G,\; \lambda,\mu \in \mathbf{C},\nonumber\\
N_{FG}=N_FN_G.
\end{eqnarray}
and
\begin{eqnarray}
[X,Y]=N_x, \nonumber\\
X N_F=N_{T_{a}F}X,\nonumber\\
Y N_F=N_{T_{-a}F}Y.
\end{eqnarray}
An involution in \(\mathcal{A}\) is defined by the following
rules:
\begin{eqnarray}
{N_F}^\star=N_{F^\star},\nonumber\\
X^\star=-Y.
\end{eqnarray}
\indent We can embed \(U(\mathfrak{sl}(2,\mathbf{C}))\), the
universal enveloping algebra of \(\mathfrak{sl}(2,\mathbf{C})\)
into \(\mathcal{A}\) if we identify each element \(X^nY^mH^k\)
from \(U(\mathfrak{sl}(2,\mathbf{C}))\) with the element \(X^nY^m
N_{x^k}\) from \(\mathcal{A}\).\newline
 \indent We will see below that \(\mathcal{A}\) has enough much
 representation, so \(\mathcal{N}\) is isomorphic to
 \(\mathcal{P}\).
 \indent Denote by \(H\) the element \(N_x\), \(H=N_x\).\newline
 \indent There
exists one-parameter group of automorphism \(U_t\;,t\in
\mathbf{C}\) of \(\mathcal{A}\) acts on generators as follows
\begin{eqnarray}
U_t(X)=e^{it}X,\;U_t(Y)=e^{-it}Y,\;U_t(N_F)=N_F.
\end{eqnarray}
\indent  The \(\star\)-subalgebra of \(\mathcal{A}\) generated by
all elements of the form \(N_F\) where \(F \in \mathcal{P}\), is
called the Cartan subalgebra and is denoted by \(\mathcal{N}\).
\newline \indent\textbf{Proposition.} The following expression
 \begin{equation}
\rho(N_F)=\int \limits_{-\infty}^{+\infty} d\sigma(x)F(x)
\end{equation}
define a state \(\rho\) on  \(\mathcal{N}\). Here \(d\sigma\) is
an arbitrary probability measure on real line which decrease
faster than any inverse polynomial. Conversely, for any state
\(\rho\) on
 \(\mathcal{N}\) there exists probability measure \(d\sigma\) on line,
which decrease faster than any inverse polynomial at infinity such
that for all \(F \in \mathcal{P}\) (20) holds.\newline
 \indent \textbf{Proof.} Let \(\varphi\) be a positive functional on the space
 \(C(\mathbf{R})\) of bounded continuous functions on
 \(\mathbf{R}\). It follows from the Riesz --- Markov theorem,
 that there exists a nonnegative measure \(d\mu\), such
 that

 a) \(\int \limits_{-\infty}^{+\infty} d\mu< \infty\),

 b) \( \varphi(g)=\int \limits_{-\infty}^{+\infty} g(x)d\mu(x)\)
 \newline
 for all continuous functions \(g(x)\) on \(\mathbf{R}\) such
 that \(g(x)\rightarrow 0\) as \(x \rightarrow \pm \infty\).

 Let us consider functionals \(\psi\) on \(C(\mathbf{R})\),
\( n=1,2,3...\),
 defined as follows
 \begin{eqnarray}
 \psi(F)=\rho(N_{(1+x^2)^{n}}N_F).
 \end{eqnarray}
 So there exist the nonnegative measures \(d\mu_n\), \(\int d\mu_n <
 \infty\) such that
 \begin{eqnarray}
 \rho((1+H^2)^n N_F)=\int \limits_{-\infty}^{+\infty} F(x)d\mu_n(x)
 \end{eqnarray}
 for all continuous function \(F(x)\) such that \(F(x)\rightarrow
 0\) as \(x\rightarrow \pm \infty\).

 So for any functions \(F(x)\) such that
\begin{eqnarray}
|F(x)|\leq C(1+x^2)^{n-1}
\end{eqnarray}
for some constant C, we have
 \begin{eqnarray}
 \rho(N_F)=\int \limits_{-\infty}^{+\infty} d\sigma_n(x) F(x),
 \end{eqnarray}
 where
\begin{eqnarray}
d\sigma_n=\frac{d\mu_n}{(1+x^2)^n}. \label{decr}
\end{eqnarray}
It is easy to see that \(d\sigma_n\) does not depend of \(n\). It
follows from the representation (\ref{decr}) that \(d\sigma_n\)
tends to zero faster then any inverse polynomial at infinity. The
proposition is proved.\newline
 \indent The proof uses the Riesz -- Markov theorem. Note that
 each positive linear functional on the space of continuous function on compact Hausdorff
 space is continuous.\newline
\indent Let us define characteristic functional of \(\rho\)
\(\chi_\rho(t)\) by the following formula
\begin{eqnarray}
\chi_\rho(t)=\rho(N_{e^{itx}}).
\end{eqnarray}
 \indent \textbf{Theorem 1.} A state \(\rho\) on \(\mathcal{N}\)
extends to a KMS state on \(\mathcal{A}\) with respect the pair
\(\{\beta, U_t\}\) (\(\beta>0\)) if and only if its characteristic
functional \(\chi_\rho(t)\) has the form
\begin{equation}
\chi_\rho(t)=m_1+m_2\frac{1-e^{-\beta}}{1-e^{-\beta+2it}}\int
\limits_{+0}^{+\infty} d\sigma(\lambda)e^{it\lambda} \label{22}
\end{equation}
for some probability measure \(\sigma\) on \((0,+\infty)\) which
decrease faster than any inverse polynomial. Here \(m_1\), \(m_2\)
are arbitrary real numbers such that \(m_1\geq0\), \(m_2\geq0\),
\(m_1+m_2=1\). If an extension exists then it is unique.
\newline
\section{Beginning of the proof.}
Let us show that the part "if" of the theorem holds. In order to
construct \(\rho\) we will investigate irreducible representations
of Lee algebra \(\mathfrak{sl}(2,\mathbf{C})\), or more precisely
modules over \(\mathcal{A}\). All irreducible representations of
\(\mathfrak{sl}(2,\mathbf{C})\) with involution (14) have
classified in [6] (see also [8]). Unitary representations of the
Lee group \(SL(2,\mathbf{C})\) have studied in [9]. We extend this
construction to the case of the algebra \(\mathcal{A}\)
 \newline
 \indent \textbf{Definition.} Let  \(\lambda\) be a real positive number. \(V_\lambda\) is a module over
 \(\mathcal{A}\) spanned on vectors \(|\lambda,n\rangle\),
 \(n=0,1,...\)
 defined by the following representation \(\;\widehat{}\;\)
 of generators
 \(X,Y,N_F\) on \(\{|\lambda,n\rangle\}\)
\begin{eqnarray}
\widehat{Y}|\lambda,p+1\rangle=-(\lambda+p)|\lambda,p\rangle,\nonumber\\
\widehat{X}|\lambda,p\rangle=(p+1)|\lambda,p+1\rangle,\nonumber\\
\widehat{Y}|\lambda,0\rangle=0,\nonumber\\
 \widehat{N}_F|\lambda,p\rangle=F(\lambda+2p)|\lambda,p\rangle,
 \nonumber\\
 p=0,1,...
\end{eqnarray}

 \textbf{Lemma 1.} For each \(\lambda>0\) there exists an unique scalar product on \(V_\lambda\)
  (defined up to arbitrary positive multiplier  ) such that
  \(\widehat{X}=-\widehat{Y}^\star\), \(\widehat{N}^\star_F=\widehat{N_F}\) with respect to this scalar product.
  \newline
 \indent Proof of this lemma is standard, see[6]\newline
 \indent \textbf{Definition.} \(V_0\) is a module over
 \(\mathcal{A}\) spanned on vector
 \(|0,0\rangle\) such that
 \begin{equation}
 \langle 0,0|0,0\rangle=1,
 \end{equation}
 defined by the following reprentation \(\;\hat{}\;\)
 of generators
 \(X,Y,N_F\) on \(|0,0\rangle\)
 \begin{eqnarray}
 \widehat{X}|0,0\rangle=\widehat{Y}|0,0\rangle=\widehat{N}_F|0,0\rangle=0.
\end{eqnarray}
\indent \textbf{KMS states \(\rho_\lambda\).} Let \(\lambda \in
\mathbf{R},\;\lambda>0\). Consider the completion
\(\bar{V}_\lambda\) of module \(V_\lambda\) with respect to a
scalar product, defined in the previous Lemma. Consider the trace
class operator \(\rho_\lambda=\frac{e^{-\beta\frac{\tilde{H}
}{2}}}{Z}\) acting in \(\bar{V}_\lambda\),  where \(\tilde{H}\) is
an unique self-adjoint extension of \(H\) from \(V_\lambda\) and
\(Z={\rm tr \mit} \{e^{-\beta\frac{\tilde{H}}{2}}\}\). Operator
\(H\) is essentially self-adjoint because \(V_\lambda\) contains
the basis of eigenvectors of \(H\), see for example [10]\newline
 Define the state \(\rho_\lambda\) on \(\mathcal{A}\) by the following
 formula
\begin{eqnarray}
\rho_\lambda(a)=\lim_{\mu\rightarrow+\infty}{\rm tr \mit}  \{{
\hat{a} E_\mu\rho}\},
\end{eqnarray}
where \(\{E_\mu\}\) is a spectral family of \(\tilde{H}\). It is
easy to proove, that this expression is well defined.\newline
 \indent \textbf{Lemma 2.} The following equality holds
\begin{eqnarray}
\rho_\lambda(N_{e^{itx}})=e^{it\lambda}\frac{1-e^{-\beta}}{1-e^{(2it-\beta)}}.\label{32}
\end{eqnarray}
  \indent\textbf{Proof.} Direct calculation.\newline
Consider a state \(\rho_0\) on \(\mathcal{A}\) defined by the
formula,
\begin{eqnarray}
\rho_0(X^mY^nN_F)=\delta_{m,0}\delta_{n,0}F(0), \nonumber\\
 n,m=0,1,..., \label{33}
\end{eqnarray}
and define a state \(\rho\) on \(\mathcal{A}\) of the form
\begin{eqnarray}
\rho(a)=m_1\rho_0(a)+m_2 \int
\limits_{+0}^{+\infty}d\sigma(\lambda)
\rho_\lambda(a), \nonumber\\
m_1\geq0,m_2\geq0,\nonumber\\
m_1+m_2=1,
\end{eqnarray}
where \(d\sigma\) is a probability measure, which decreases faster
than any inverse polynomial. By using the definition of
\(V_\lambda\) and the scalar product on it one can see that
\(\forall a \in \mathcal{A}\) \(\rho_\lambda(a)\) increase slowly
than some polynomial at infinity. So the integral in the right
hand side exists. It follows from lemma 2 that the characteristic
functional of the restriction of \(\rho\) on \(\mathcal{N}\) has a
needed form (\ref{22}). So the part "if" is proved. \newline

 \section{Decomposition of the state \(\rho\) into the direct integral.}
 Now we begin to prove the part "only if". Let \(\rho\) be an
 KMS functional on \(\mathcal{A}\).

Let us make the GNS construction for the state \(\rho\). We get a
Hilbert space \(\mathcal{H}\), the dense subspace \(\mathcal{D}\),
the representation \(\hat{{}}\) of \(\mathcal{A}\) by means
operators, acting from \(\mathcal{D}\) to \(\mathcal{D}\), cyclic
vector \(|\Omega\rangle\in \mathcal{D}\) i.e. the vector such that
\(\hat{\mathcal{A}}|\Omega\rangle=\mathcal{D}\). We get also
\(\rho(a)=\langle\Omega|\hat{a}|\Omega\rangle\). For each \(a \in
\mathcal{D}\) by \(|a\rangle\) denote the vector
\(\hat{a}|\Omega\rangle\).
\newline \indent \indent\textbf{Lemma 3.} There exists an unique
projector-valued measure \(dE\) in \(\mathcal{H}\) such that for
all \(|f\rangle,|g\rangle \in \mathcal{D}\) and a continuous
function \(F(\lambda)\), which increase slowly than some
polynomial at infinity.
\begin{eqnarray}
\langle f|\widehat{N}_F|g\rangle=\int
\limits_{-\infty}^{+\infty}F(\lambda)\langle
f|dE(\lambda)|g\rangle,
\end{eqnarray}
and \(\langle f|dE(\lambda)|g\rangle\) decrease faster than any
inverse polynomial.\newline
 \indent \textbf{Proof.} The functional \(\rho\) is positive. So
 for all \(f \in \mathcal{D}\) the functional
 \(F\rightarrow\langle f|\hat{N}_F|f\rangle\) is positive.
 Therefore there exists a Borelian measure \(d\mu_{f,f}\) which
 decrease faster then any inverse polynomial such that
 \begin{eqnarray}
\langle f|\hat{N}_F|f\rangle=\int
\limits_{-\infty}^{+\infty}F(\lambda)d\mu_{f,f}(\lambda).
\end{eqnarray}
 Using polarization identity we can find the measure
\(d\mu_{f,g}\), such that
 \begin{eqnarray}
\langle f|\hat{N}_F|g\rangle=\int
\limits_{-\infty}^{+\infty}F(\lambda)d\mu_{f,g}(\lambda).\label{meas}
\end{eqnarray}
The measure \(d\mu_{f,g}\) is a linear functional of \(g\) and an
antilinear functional of \(f\). Now, for each bounded Borelian
function \(F\) define the following sesqulinear form
 \begin{eqnarray}
 \mathcal{N}_F(f,g):=\int
 \limits_{-\infty}^{+\infty}F(\lambda)d\mu_{f,g}.
\end{eqnarray}
It follows from this representation that for each bounded Borelian
function \(F\) there exists bounded operator in \(\mathcal{H}\)
which we denote by \(\hat{N}_F\) such that
\begin{eqnarray}
 \mathcal{N}(f,g)=\langle f|\hat{N}_F|g\rangle
 \end{eqnarray}
 and
\begin{eqnarray}
\|\hat{N}_F\|\leq 4 \sup \limits_{x \in\mathbf{R}}|F(x)|.
\end{eqnarray}
Now for each \(f,g \in \mathcal{H}\) (not necessary in
\(\mathcal{D}\)) we can define the measure \(d\mu_{f,g}\) by the
following formula
\begin{eqnarray}
\mu_{f,g}(B)=\langle f|\hat{N}_{\chi_B}|g\rangle,
\end{eqnarray}
where \(\chi_B\) is an indicator of Borelian set \(B\).

Let us prove that the measure \(d\mu_{f,g}\) is
\(\sigma\)-additive measure. Let \(B_n,\;n=1,2,...\) the sequence
of Borelian sets, such that
\begin{eqnarray}
B_1\subset B_2 \subset...
\end{eqnarray}
Let \(B=\bigcup \limits_n B_n\). \(\forall\varepsilon>0\), \(f,g
\in \mathcal{H}\) there exist \(N>0\),
\(\tilde{f},\tilde{g}\in\mathcal{D}\) such that
\begin{eqnarray}
|\langle f|\hat{N}_{\chi_{B_n}}|g\rangle\ -\langle
\tilde{f}|\hat{N}_{\chi_{B_n}}|\tilde{g}\rangle|<\varepsilon,\nonumber\\
|\langle \tilde{f}|\hat{N}_{\chi_{B}}|\tilde{g}\rangle -\langle
{f}|\hat{N}_{\chi_{B}}|{g}\rangle|<\varepsilon,\nonumber\\
|\langle \tilde{f}|\hat{N}_{\chi_{B_n}}|\tilde{g}\rangle-\langle
\tilde{f}|\hat{N}_{\chi_{B}}|\tilde{g}\rangle|<\varepsilon,
\end{eqnarray}
if \(n> N\). We have \(|\langle
f|\hat{N}_{\chi_{B_n}}|g\rangle-\langle\tilde{f}|\hat{N}_{\chi_{B}}|\tilde{g}\rangle|<3\varepsilon
\), therefore \(\mu_{f,g}(B_n) \rightarrow \mu_{f,g}(B)\), and
\(\mu\) is \(\sigma\)-additive. Now using approximation of
Borelian function \(F\) by simple function we can prove that
(\ref{meas}) is valid for all \(f,g \in \mathcal{H}\) and bounded
Borelian function \(F\).

Let \(F\), \(G\) --- be continuous functions with compact support.
We have
\begin{eqnarray}
\langle f|\hat{N}_F\hat{N}_G|g\rangle=\langle
f|\hat{N}_{FG}|g\rangle. \label{mul}
 \end{eqnarray}
  We have proved the representation
(\ref{meas}) for all \(f,g \in \mathcal{H}\) and bounded Borelian
function \(F\). From this fact it follows that formula (\ref{mul})
is valid for all bounded Borelian functions \(F,G\). So we have
constructed projector-valued measure \(E(B)=\widehat{N}_{\chi_B}\)
such that for all \(f,g \in \mathcal{D}\) and \(F(\lambda) \in
\mathcal{P}\)
\begin{eqnarray}
\langle f|\hat{N}_F|g\rangle=\int
\limits_{-\infty}^{+\infty}F(\lambda) \langle
f|dE_\lambda|g\rangle.
\end{eqnarray}
The lemma is proved
\newline
 \indent\textbf{Remark.} For any function \(F\) which increase slowly than
some polynomial the following spectral decomposition
\(\widetilde{{F}(H)}=\int \limits_{-\infty}^{+\infty}F(\lambda)
 dE(\lambda)\) define a normal operator in \(\mathcal{H}\), which
extends \(\widehat{N}_F\).\newline
 \indent
\textbf{Lemma 4.} \({\rm Sp \mit} (\tilde{H}) \), the spectrum of
\(\tilde{H}\) lies in \([0,+\infty)\).\newline
 \indent \textbf{Proof}.
Let \(f\) be a continuous function with a compact support.
 Let us compute \(-\rho(X {{N}_f}Y)\), using the KMS property of the state
 \(\rho\).
 \begin{eqnarray}
 -\rho(X {N}_fY)=-\rho(\hat{N}_fYU_{i\beta}(X)).
 \end{eqnarray}
 But \(U_{i\beta}(X)=e^{-\beta}X\), therefore
\begin{eqnarray}
-\rho(X {N}_{f}Y)=e^{-\beta}\rho({N}_f{N}_x)-e^{-\beta}\rho(X
{N}_{T_{-2}f}Y),
\end{eqnarray}
or
\begin{eqnarray}
-\rho(X {N}_{T_{-2}f}Y)=-e^{\beta}\rho(X {N}_fY)-\rho({N}_f{N}_x).
\end{eqnarray}
Substituting \(f(x)\) for \(f(x+2k)\) in previous equality we get
\begin{eqnarray}
-\rho(X {N}_{T_{-2-2k}f}Y)=-e^{\beta}\rho(X {N}_{T_{-2k}f} Y)-\rho({N}_{T_{-2k}f}{N}_x),\nonumber\\
k=0,1,2...
\end{eqnarray}
Let \(f\) be a continuous function with a compact support \(\rm
supp \mit f\subset(-\infty,0)\). Suppose that
 \(\rho({N}_f)=0\) (this fact will be proven below).
Then
\begin{eqnarray}
\forall\; |a\rangle, |b\rangle \in \mathcal{D}
 \langle a|\widehat{N}_f|b\rangle=0.
\end{eqnarray}
 Indeed
\begin{eqnarray}
\langle a|\widehat{N}_f|b\rangle=\rho(a^\star\widehat{N}_f
b)=\rho(U_{-i\beta}(b)a^\star\widehat{N}_f).
\end{eqnarray}

Using Schwarz inequality
\begin{eqnarray}
|\langle a|\widehat{N}_f|b\rangle|\leq|\rho(
(U_{-i\beta}(b))a^\star a
 U_{-i\beta}(b))|^{\frac{1}{2}}\rho({N}_f^\star
{N}_f )^{\frac{1}{2}}=0.
\end{eqnarray}
It follows from (35) that \({\rm Sp \mit}
(\tilde{H})\subset[0,+\infty)\). So suppose that there exists a
segment \([a,b]\subset (-\infty,0)\) such that
\(\langle\Omega|E([a,b])|\Omega\rangle))\neq 0\). Thus there
exists a continuous function \(f\) with compact support \(\rm supp
\mit f\subset (-\infty,0)\), \(f\geq 0\)  such that
 \begin{equation}
 \rho({N}_f{N}_x)<0.
 \end{equation}
 It follows from positivity of \(f\) that:
\begin{eqnarray}
\rho({N}_{T_{-2k}f}{N}_x)\leq0, k=1,2,3...\nonumber
\end{eqnarray}
Moreover, it follows from positivity \(f\) that:
\begin{eqnarray}
-\rho(X{N}_{T_{-2k}f} Y)\geq0, k=1,2,3... \nonumber
\end{eqnarray}
So, we get
\begin{eqnarray}
-\rho(X{N}_{T_{-2k-2}f}  Y)\geq -e^{\beta}\rho(X {N}_{T_{-2k}f}Y),\nonumber\\
k=1,2...\nonumber\\
-\rho(X{N}_{T_{-2}f}Y))>0.
\end{eqnarray}
From (39) we get that \(-\rho(X{N}_{T_{-2k}f}Y)\) tends
exponentially to infinity then \(k\rightarrow+\infty\), but
\(-\sum \limits_{k=1}^{+\infty}\rho(X{N}_{T_{-2k}f}Y)<+\infty\).
This contradiction concludes the proof .\newline
 \indent \textbf{Decomposition of the state \(\rho\) into the direct integral.}\newline
 Let \(\mathbf{C'}\) be the ring of all finite linear combination
 of elements of the form \(e^{n\beta}e^{i\pi r}\), \(r
 \in\mathbf{Q}\), \(n \in \mathbf{Z}\) with rational coefficients. It is obvious that this
 ring contains only countable number of elements.\newline
 \indent \textbf{Definition.} Let \(A=\{\eta_i\}\), \(i\in \mathbf{Z}\)
 some countable set of continuous function with compact support.
 Let \(C\) be a set of functions, which consist of all elements of the form
 \begin{eqnarray}
 \prod
 \limits_{i=1}^{n}\eta_{i_1}(x-2l_1)...\eta_{i_n}(x-2l_n),\nonumber\\
 n=0,1,2...,
 l_i \in \mathbf{Z}.
 \end{eqnarray}
 It is obvious that \(C\) is countable. Let us consider the set of
 all elements of the form:
\begin{equation}
P(X,Y,H)N_f N_{e^{i\pi r x}},\; r \in \mathbf{Q},\nonumber\\
f(x) \in C,
\end{equation}
where \(P(X,Y,H)\) is a polynomial on its arguments with
 coefficient from \(\mathbf{C}'\). This set is countable. The
 algebra \(\mathcal{A}^\star\) over \(\mathbf{C}'\), by definition, consists of all linear combination of elements of the
 form (41) with rational coefficients.\newline
\indent Let \(U:=N_{e^{i\pi {x}}}\) be an unitary element from
 \(\mathcal{A}^\star\). It is clear that this element is a central
 element of \(\mathcal{A}^\star\). We get that \(\tilde{U}=e^{i\pi \tilde{H}}\) be an unitary
 operator, which acts in \(\mathcal{H}\).
 Let
 \begin{equation}
 \tilde{U}=\int \limits_{0}^{2\pi} e^{i\varphi}dP_\varphi
 \end{equation}
be its spectral decomposition. \(\forall a \in
\mathcal{A}^{\star}\) we put by definition \(d\eta
(a)=\langle\Omega|dP_\varphi \hat{a}|\Omega\rangle\), and
\(d\mu=\langle\Omega|dP_\varphi|\Omega\rangle\). \newline \indent
\textbf{Lemma 5.} For almost all
 \(\varphi \in [0,2\pi)\) there exists a unique KMS-state
 \(\tau_\varphi\) on \(\mathcal{A}^\star\)
 such that
\begin{eqnarray}
d\eta(a)=\tau_\varphi(a) d\mu(a).
\end{eqnarray}
\textbf{Proof.} Obviously we have
\begin{equation}
P_{[\varphi_1,\varphi_2]}=\sum \limits_{n=-\infty}^{n=+\infty}
E_{[\frac{\varphi_1}{\pi}+2n,\frac{\varphi_2}{\pi}+2n]}.
\end{equation}
Let us consider the following measures
\begin{eqnarray}
d\eta(a)=\langle\Omega|(dP_\varphi\hat{a})|\Omega\rangle,\nonumber\\
d\mu=\langle\Omega|(dP_\varphi)|\Omega\rangle.
\end{eqnarray}
Let us prove that the measure \(d\eta(a)\) is absolutely
continuous measure with respect the measure \(d\mu\). Indeed let
\(B\) be a Borelian set such that \(\mu(B)=0\). We have
\begin{eqnarray}
|\eta(a)(B)|=
|\langle\Omega|P(B)\hat{a})|\Omega\rangle|\leq\langle\Omega|(\hat{a}\hat{a}^\star)|\Omega\rangle^{\frac{1}{2}}
\langle\Omega|(P(B)P(B))|\Omega\rangle^{\frac{1}{2}}
=\nonumber\\
=\rho(aa^\star)^{\frac{1}{2}}\langle\Omega|P(B)|\Omega\rangle^{\frac{1}{2}}=0.
\end{eqnarray}
So, by using the Radon --- Nickodym theorem we see that there
exists the function \(\tau(a)(\varphi)\) such that
\begin{equation}
\tau(a)(\varphi)d\mu(\varphi)=d\eta(a).
\end{equation}
We have
\begin{equation}
\tau(\lambda a_1+\mu
a_2)(\varphi)=\lambda\tau(a_1)(\varphi)+\mu\tau(a_2)(\varphi).
\end{equation}
The last equality is valid almost everywhere.

Let us prove that \(\tau(a^\star a)(\varphi)\geq 0\) for almost
all \(\varphi\) and all \(a \in \mathcal{A}^\star\). Let
\(P(e^{i\pi \varphi})\) be an arbitrary positive trigonometric
polynomial. According to the Riesz theorem we find that there
exists a trigonometric polynomial  \(Q(e^{i\pi \varphi})\) such
that
\begin{equation}
P(e^{i\varphi})=Q^\star(e^{i\varphi})Q(e^{i\varphi}),\nonumber\\
 P(N_{e^{i\pi x}})=Q^\star(N_{e^{i\pi x}})Q(N_{e^{i\pi x}}).
\end{equation}
Using the spectral decomposition of \(e^{i\pi \tilde{H}}\) we see
\begin{eqnarray}
\int \limits_0^{2\pi}
\tau(aa^\star)(\varphi)P(e^{i\varphi})d\mu(\varphi)=\int
\limits_0^{2\pi} \langle\Omega|\widehat{aa^\star}
dP(\varphi)|\Omega\rangle P(e^{i\varphi})=\rho(aa^\star
P(N_{e^{i\pi x}}))\geq0.
\end{eqnarray}
So for almost all \(\varphi\) and all \(a \in\mathcal{A}^\star\)
\(\tau(a^\star a)(\varphi)\geq0\). Note that we use the fact that
\(\mathcal{A}^\star\) contains only countable number of elements.

Let us prove that \(\tau(\varphi)\) is a KMS-functional. Note that
\(\rho(\cdot P(N_{e^{i\pi x}}))\) is a KMS-functional for an
arbitrary trigonometric polynomial \( P(N_{e^{i\pi \varphi}})\)
i.e.
\begin{equation}
\rho(ABP(N_{e^{i\pi x}}))=\rho(BU_{i\beta}(A)P(N_{e^{i\pi x}})).
\end{equation}
Using spectral decomposition for \(e^{i\pi \tilde{H}}\) we find:
\begin{equation}
\int \limits_0^{2\pi}
\tau(AB)(\varphi)P(e^{i\varphi})d\mu(\varphi)= \int
\limits_0^{2\pi}
\tau(BU_{i\beta}(A))(\varphi)P(e^{i\varphi})d\mu(\varphi).
\end{equation}
 \(P(e^{i\varphi})\) is an arbitrary trigonometric polynomial. So
 for almost all \(\varphi\) and  all \(A, B \in \mathcal{A}^\star\)
 we find:
\begin{equation}
 \tau(AB)(\varphi)=\tau(BU_{i\beta}(A))(\varphi).
\end{equation}
The lemma is proved.\newline
 The proof of this lemma is like to the proof of
 the von Neumann spectral theorem [11].  \newline
 Let us make now for all \(\varphi\) from the previous lemma the GNS
 construction for \(\tau_\varphi\). We get:\newline
 a) The Hilbert space \(\mathcal{H}_\varphi\),\newline
 b) The dense subspace \(\mathcal{D}_\varphi\) over the ring \(\mathbf{C'}\).\newline
 c) The representation \(\hat{{}}\) of \(\mathcal{A}^\star\) in
 \(\mathcal{H}_\varphi\) by means
 \(\mathbf{C'}\)-linear operator, acting from \(\mathcal{D}_\varphi\) to \(\mathcal{D}_\varphi\).\newline
 d) The vector \( |\Omega_\varphi\rangle\in \mathcal{D}_\varphi\)
 such that
 \(\mathcal{A}^\star|\Omega_\varphi\rangle=\mathcal{D}_\varphi\).
 \newline
 \indent \textbf{Definition.} The algebra \(\mathcal{A}^{\star\star}\) is an algebra
 generated by all elements of the form
 \begin{eqnarray}
 P(X,Y,N_x)N_fN_{e^{i\pi rx}}, \nonumber\\
 r \in \mathbf{}{Q}.
 \end{eqnarray}
 Here \(P(X,Y,N_x)\) is a polynomial, and \(f\) is an element of
 \(C\) of the form
 \begin{eqnarray}
 f(x)=\prod \limits_{i=1}^m \eta_i(x-2k_i)\nonumber\\
 m=1,2...
 \end{eqnarray}
 Let  \(\mathcal{D'}_\varphi\) by definition be subspace of
 \(\mathcal{D}_\varphi\) of the form
 \(\mathcal{D'}_\varphi={\widehat{A^{\star\star}}}|\Omega_\varphi\rangle\).\newline
  \indent \textbf{Lemma 6.} We can chose the set \(A\) such that  \(\mathcal{D'}_\varphi\) is a dense
  subspace of \(\mathcal{D}_\varphi\) (for almost all \(\varphi\)).\newline
\indent \textbf{Proof.} Let \(a \in \mathcal{A}^{\star}\) and be a
\({\eta}_n(x) \in B \) sequence of real-valued continuous function
such that \( \rm supp \mit  \eta_n(x) \in [-2n,2n]\) and \(0\leq
\eta_n(x)\leq 1\), \(\eta_n(x)|_{[-n,n]}=1\). Let us prove that
\(\mathcal{D'}_\varphi\) is a dense subset in
\(\mathcal{D}_\varphi\). We have:
 \begin{eqnarray}
 \|(\hat{a}\hat{N}_{\eta_n(x)}-\hat{a})|\Omega_\varphi\rangle\|=\tau(a^\star a(N_{\eta_n(x)}-1)^2)(\varphi)\leq
 \nonumber\\
\leq \tau((a^\star
a)^2)^{1/2}\tau((N_{\eta_n(x)}-1)^4)^{1/2}(\varphi). \label{ineq}
\end{eqnarray}
But \(\rho((N_{\eta_n(x)}-1)^4)^{1/2}\rightarrow 0\). So there
exists subsequence \(\eta'_k\) of the sequence \(\eta_n(x)\) such
that
 \begin{eqnarray}
\rho((N_{\eta'_n(x)}-1)^4)\leq \frac{1}{2^n}.
\end{eqnarray}
Therefore the following series
\begin{eqnarray}
\sum \limits_{n=1}^\infty \int \tau((
N_{\eta'_n(x)}-1)^4)(\varphi)d\mu(\varphi)
\end{eqnarray}
converges. So by using B. Levi theorem we find that
\(\tau((\eta'_n(H)-1)^4)^{1/2}(\varphi)\rightarrow 0\) for almost
all \(\varphi\). This fact and inequality (\ref{ineq}) implies
that \(\|(a N_{\eta'_n(x)}-a)|\Omega_\varphi\rangle\|\rightarrow
0\). The lemma is proved.\newline
 \indent The following lemma holds. \newline
 \indent \textbf{Lemma 7.}
 For all \(\varphi\) from lemma 5 there exists the spectral
 family \(F_{\pi^{-1}\varphi+2n}\) in \(\mathcal{H}_\varphi\):
\begin{eqnarray}
F_{\pi^{-1}\varphi+2n}F_{\pi^{-1}\varphi+2m}=F_{\pi^{-1}\varphi+2n}\delta_{n,m},\nonumber\\
\sum \limits_{n=0}^{+\infty} F_{\pi^{-1}\varphi+2n}=I,\nonumber\\
 n,\;m=0,1,2....\nonumber\\.
\end{eqnarray}
such that
\begin{eqnarray}
\langle f|\widehat{N}_G|g\rangle=\sum \limits_{n=0}^{+\infty}
G(\pi^{-1}\varphi+2n) \langle f| F_{\pi^{-1}\varphi+2n}|g\rangle.
\end{eqnarray}
For all \(|f\rangle,|g\rangle \in \mathcal{D}'_\varphi\).
Moreover, self-adjoint operator, which acts in
\(\mathcal{H}_\varphi\) defined by its spectral decomposition
\begin{eqnarray}
\tilde{H}_\varphi=\sum \limits_{n=0}^{+\infty}
(\pi^{-1}\varphi+2n) F_{\pi^{-1}\varphi+2n}
\end{eqnarray}
is positive.\newline
 \indent \textbf{Proof.} Let \(K\) be a smooth function with
 compact support such that \({N}_K \in \mathcal{A}^{\star \star}\).
 Let \(N \in \mathbf{Z}^+\) be a number such that \( \rm supp \mit
 K \subset [-N,N]\).

 Note that the measure
 \begin{eqnarray}
 \rho(f^{\star}dE_{{\pi}^{-1}\varphi+2n }g),
 \end{eqnarray}
 where \(f,g \in \mathcal{A}^{\star\star}\) is an absolutely continuous measure with respect the measure
 \(d\mu\) because
\begin{eqnarray}
dP_\varphi=\sum \limits_{n=-\infty}^{n=\infty}dE_\varphi.
 \end{eqnarray}
So there exists the functions \(\psi_n(\varphi)[f,g] \in
L_1(d\mu)\) such that
 \begin{eqnarray}
\rho(f^{\star}dE_{{\pi}^{-1}\varphi+2n
}g)=\psi_n(\varphi)[f,g]d\mu(\varphi).
\end{eqnarray}
Note that for almost all \(\varphi\) \(\psi_n(\varphi)[f,g]\) is a
positive sesqulinear form and
\begin{eqnarray}
\tau(\varphi)(f^\star g)=\sum
\limits_{n=-\infty}^{n=+\infty}\psi_n(\varphi)[f,g]. \label{LL}
\end{eqnarray}
 Let \(G(e^{i\pi\varphi})\) be an arbitrary trigonometric
 polynomial. We have
\begin{eqnarray}
\rho(f^\star N_K N_{G(e^{i\pi\varphi})} g) =\sum
\limits_{n=-N}^{n=N}\int
K({\pi}^{-1}\varphi+2n)\psi_n(\varphi)[f,g]G(e^{i\pi \varphi})
d\mu(\varphi)=\nonumber\\
=\int \tau(f^\star N_K g)(\varphi)G(e^{i\pi \varphi}) d\varphi.
\end{eqnarray}
 So we have
\begin{eqnarray}
\tau(f^\star N_K g)(\varphi)=\sum \limits_{n=-N}^{n=N}
K({\pi}^{-1}\varphi+2n)\psi_n(\varphi)[f,g]. \label{KK}
\end{eqnarray}
Suppose that the function \(K_n\) has a support in a small
neighborhood of the point \({\pi}^{-1}\varphi+2n\), and
\(K_n({\pi}^{-1}\varphi+2n)=1\). We have
\begin{eqnarray}
\tau(f^\star N_{K_n} g)(\varphi)=\psi_n(f,g). \label{SS}
\end{eqnarray}
It follows from this identity that \(\hat{N}_K\) is self-adjoint
bounded operator in \(\mathcal{H}_\varphi\). It follows from
(\ref{SS}) that
\begin{eqnarray}
\tau(f^\star N_{K_n} N_{K_n} g)(\varphi)=\psi_n(f,g) \label{PP}.
\end{eqnarray}
It follows from (\ref{PP}) that \(\widehat{N}_{K_n}\) is a
projector in \(\mathcal{H}_\varphi\). Let
\(F_{{\pi}^{-1}\varphi+2n}=\widehat{N}_{K_n}\). One can easily
proof using (\ref{KK}) that
\(F_{{\pi}^{-1}\varphi+2n}F_{{\pi}^{-1}\varphi+2m}=0\) if \(n\neq
m\). The fact, that \(F_{{\pi}^{-1}\varphi+2n}=0\) if \(n<0\)
follows from the positivity of \(\tilde{H}\) in \(\mathcal{H}\).
It follows from (\ref{LL})
\begin{eqnarray}
\sum \limits_{n=-\infty}^{n=+\infty} F_{{\pi}^{-1}\varphi+2n}=1.
\end{eqnarray}

\section{Decomposition \(\mathcal{H}_\varphi\) into the sum of irreducible components.}
The following lemma holds. \newline
 \indent \textbf{Lemma 8.} \( \mathcal{H}_\varphi\) can be decomposed into the direct sum of
subspaces
 \(\mathcal{H}_\varphi^{k}\), \(k=0,1,2,...\)
\begin{equation}
 \mathcal{H}_\varphi=\bigoplus \limits_{k=0}^{\infty} \mathcal{H}_\varphi^{k}
\end{equation}
such that\newline
 a) For all \(m=0,1,2...\) operators
\(\hat{X},\hat{Y},\widehat{N}_F \) (\({N}_F\in A^{\star\star}\))
extends by continuity to bounded operators from  \(\rm Ran \mit
F_{\pi^{-1}\varphi+2m}\) to \(\mathcal{H}_\varphi\). These
extensions we will also denote by
 \(\hat{X},\hat{Y},\widehat{N}_F)\).\newline
 c) The following subspaces
\begin{equation}
\bigcup \limits_m \mathcal{H}_{\varphi}^n \cap \{\rm Ran \mit
F_{\pi^{-1}\varphi}\oplus...\oplus \rm Ran \mit
F_{\pi^{-1}\varphi+2m}\}
\end{equation}
are invariant under the action of the operators \(\hat{X},
\hat{Y}, \widehat{N_F}\). \newline
 d)
 \begin{eqnarray}
 \mathcal{H}_\varphi^n\cap \rm Ran \mit F_{\pi^{-1}\varphi+2k}=0 \; \rm if \mit \; k<n, \nonumber\\
 \hat{X}^l(\mathcal{H}_\varphi^n\cap \rm Ran \mit F_{\pi^{-1}\varphi+2k})=\mathcal{H}_\varphi^{n}\cap \rm
 Ran\mit
 F_{\pi^{-1}\varphi+2k+2l}, \; l>0,\; \; \rm if \mit \; k\geq n.
\end{eqnarray}

\textbf{Proof} We have proved that \(\tau(a)(\varphi)\) is a KMS
state on \(\mathcal{A}^{\star\star}\) for all \(\varphi\) from
\([0,2\pi)\setminus A\), where \(\mu (A)=0\). Let \(\varphi\) be
an element from \([0,2\pi)\setminus A\). Let \(n_0\) be a minimal
integer number such that \(F_{\pi^{-1} \varphi+2n_0}\neq 0\).

 \(\mathcal{D}_\varphi\) is a dense subset in
 \(\rm Ran \mit(F_{\pi^{-1}\varphi+2n_0})\).

 Let \(\eta(\lambda)\) be a smooth function such that \(\rm supp
 \mit \eta(\lambda)\) is placed at a small neighborhood of the
 point \(\pi^{-1}\varphi+2n_0\), and \(\eta(\pi^{-1}\varphi+2n_0)=1\).
 So \(\widetilde{\eta(H)}=F_{\pi^{-1}\varphi+2n_0}\). We can think
 that \(N_\eta \in \mathcal{A}^{\star\star}\).
Using the Pythagoras theorem we find that
\(D_{\pi^{-1}\varphi+2n_0}:=\mathcal{D}_\varphi\cap
 \rm Ran \mit(F_{\pi^{-1}\varphi+2n_0})\) is a dense set in
 \(\rm Ran \mit(F_{\pi^{-1}\varphi+2n_0})\).\newline
 Let us define the following operators \(\tilde{X}, \tilde{Y}\) acting in \(\widehat{\mathcal{A^{\star\star}}}D_{\pi^{-1}\varphi+2n_0}\)
 according with the following formula:
\begin{eqnarray}
\tilde{X}=\lim \limits_{n\rightarrow\infty} \widetilde{\eta_n(H)} \hat{X}, \nonumber\\
\tilde{Y}=\lim \limits_{n\rightarrow\infty} \widetilde{\eta_n(H)}
\hat{Y}, \nonumber\\ \label{lim}
\tilde{H}=\lim
\limits_{n\rightarrow\infty} \widetilde{\eta_n(H)} \hat{H}.
\end{eqnarray}
Here \(\eta_n(H) \in B\) --- is a sequence of real-valued
functions such that\newline
  a) \(0\leq\eta_n(H)\leq 1\), \newline
  b) \(\rm supp \mit \eta_n(\lambda)\subset [-2n, 2n]\), \newline
  c) \(\eta_n(H)|_{[-n,n]}=1\),\newline
  d) \(N_{\eta_n} \in \mathcal{A^{\star\star}}\).

  To define the limits (\ref{lim}) we need no any topology because
  \( \widetilde{\eta_n(H)}\hat{X}, \widetilde{\eta_n(H)} \hat{Y}\)
  become stabilize on \(\widehat{\mathcal{A^{\star\star}}}D_{\pi^{-1}\varphi+2n_0}\).
 Note that the following relation holds
  \begin{eqnarray}
[\tilde{X},\tilde{Y}]=\widetilde{N_x}, \nonumber\\
\tilde{X} \widetilde{N_F}=\widetilde{N}_{T_{a}F}\widetilde{X},\nonumber\\
\tilde{Y} \widetilde{N_F}=\widetilde{N}_{T_{-a}F}\widetilde{Y}.
\end{eqnarray}
and
\begin{eqnarray}
{N_F}^\star=N_{F^\star},\nonumber\\
X^\star=-Y.
\end{eqnarray}
Note that \(\forall n \in\mathbf{Z}\) the operators
\(\tilde{X},\tilde{Y},\tilde{H}\) are the bounded operators from
\(F_{\pi^{-1}\varphi+2n}\cap\widehat{\mathcal{A^{\star\star}}}D_{\pi^{-1}\varphi+2n_0}\)
to \(\mathcal{H}_\varphi\). The proof of this fact is similar to
derivation of the formula for scalar product on \(V_\lambda\). So
we can extend the operators \(\tilde{X},\tilde{Y},\tilde{H}\) to
the operators acting in
\(\overline{\mathcal{A}^{\star\star}D_{\pi^{-1}\varphi+2n_0}} \)
  with invariant domain:
 \(\rm Lin \mit\{\bigcup \limits_{n} \overline{\rm Ran \mit F_{\pi^{-1}\varphi+2n}\cap
\widehat{
\mathcal{A^{\star\star}}}\mathcal{D}_{\pi^{-1}\varphi+2n_0}}\}\)

It is easy to see that
\begin{eqnarray}
\overline{\widehat{\mathcal{A}^{\star\star}}\mathcal{D}_{\pi^{-1}\varphi+2n_0}}
=\overline{\bigcup \limits_n \tilde{X}^n \rm Ran \mit
F_{\pi^{-1}+2n_0}}.
\end{eqnarray}

Note that the formulas (\ref{lim}) defines operators
\(\widetilde{X},\widetilde{Y},\widetilde{H}\) acting in
\(\widehat{\mathcal{A}^{\star\star}}\rm Lin \mit \{\bigcup
\limits_n (\rm Ran \mit F_{\pi^{-1}+2n}\cap\mathcal{D})\). Denote
by \(\mathcal{H}_\varphi^{2n_0}\) the space
\(\overline{\widehat{\mathcal{A}^{\star\star}}\mathcal{D}_{\pi^{-1}\varphi+2n_0}}\).

Let us prove that for all \(n=1,2,...\) the operators
\(\tilde{X}^n\) are the bounded operators from
 \(\rm
 Ran \mit
F_{\pi^{-1}\varphi+2n_0+2}\cap\mathcal{D}_{\varphi}\) to
\(\mathcal{H}_\varphi\). Let \(\psi \in \rm Ran \mit
F_{\pi^{-1}\varphi+2n_0+2}\). We can represent \(\psi\) as a sum:
\begin{eqnarray}
\psi=f_1+f_2, \nonumber\\
f_1 \in \mathcal{D}_{\varphi}\cap \mathcal{H}_{\varphi}^{n_0} \cap
F_{\pi^{-1}\varphi+2n_0+2}, \nonumber\\
f_2 \in \mathcal{D}_{\varphi} \cap F_{\pi^{-1}\varphi+2n_0+2}.
\label{138}
\end{eqnarray}
For all \(\varepsilon>0\) we can find decomposition (\ref{138})
such that the projection of the vector \(f_2\) to the space
\(\mathcal{H}_{\varphi}^{n_0}\cap F_{\pi^{-1}\varphi+2n_0+2}\) has
a norm which is less then \(\varepsilon\). So we can think
 \(\|f_1\|\leq 2\|\psi\|\), \(\|f_2\|\leq 2\|\psi\|\).
Let us calculate \(\langle
f_2|\tilde{Y}^n\tilde{X}^n|f_2\rangle\).

\(\forall n \in \mathbf{Z}^+\) we will prove by induction there
exists constant \(C_n\) such that
\begin{eqnarray}
\langle f_2|\tilde{Y}^n\tilde{X}^n|f_2\rangle\leq C_n\parallel
f_2\parallel^2.
\end{eqnarray}
We have
\begin{eqnarray}
\langle f_2|\tilde{Y}^{n+1}\tilde{X}^{n+1}|f_2\rangle=\nonumber\\
=\langle f_2|\tilde{Y}^{n}\tilde{X}^{n+1}\tilde{Y}|f_2\rangle+
\sum \limits_{i=0}^{n+1}\langle
f_2|\tilde{Y}^{n}\tilde{X}^{i}[\tilde{Y},\tilde{X}].
\tilde{X}^{n-i}|f_2\rangle
\end{eqnarray}
The second term in the right hand side of last equality is equal
to
\begin{eqnarray}
C\langle f_2|\tilde{Y}^n\tilde{X}^n|f_2\rangle
\end{eqnarray}
for some constant \(C\) and we must to estimate the first term
\(\langle f_2|\tilde{Y}^{n}\tilde{X}^{n+1}\tilde{Y}|f_2\rangle\).
Note that \(\tilde{Y}|f_2\rangle \in \rm Ran \mit \in
F_{\pi^{-1}\varphi+2n_0}\) and there exists the constant \(C'\)
such that \(\|\tilde{Y}|f_2\rangle \|\leq\parallel f_2\parallel\).
We have proven that \(\forall n=0,1,2...\) the operators
\(\tilde{X},\tilde{Y},\tilde{H}\) are the bounded operators on
\(\rm Ran \mit
F_{\pi^{-1}\varphi+2n+2n_0}\cap\widehat{A^{\star\star}}D_{\pi^{-1}\varphi+2n_0}\),
\(n=0,1,2...\). So there exists a constant \(C''\) such that
\(\langle f_2|\tilde{Y}^{n}\tilde{X}^{n+1}\tilde{Y}|f_2\rangle\leq
C''\langle f_2| f_2\rangle\). The statement is proved.

So all the powers of \(\tilde{X}\) we can extend from \(\rm Ran
\mit F_{\pi^{-1}\varphi+2n_0+2}\cap\mathcal{D}_\varphi\) to the
\(\rm Ran \mit F_{\pi^{-1}\varphi+2n_0+2}\). It is easy to prove
as above that the operators \(\tilde{X},\tilde{Y},\tilde{H}\) are
the bounded operators on \(\tilde{X}^n(\rm Ran \mit
F_{\pi^{-1}\varphi+2n_0+2}\cap\mathcal{D}_\varphi)\). So we can
extend the operators \(\tilde{X},\tilde{Y},\tilde{H}\) to the
operators which acts in the space
\begin{eqnarray}
\rm Lin \mit \{ \bigcup \limits_n \widetilde{X}^n (\rm Ran \mit
F_{\pi^{-1}\varphi+2n_0}\oplus
 \rm Ran \mit F_{\pi^{-1}\varphi+2n_0+2})\}.
 \end{eqnarray}
 For all \(N=1,2,3...\) the restrictions of this operators to the spaces
\begin{eqnarray}
\rm Lin \mit\{ \bigcup \limits_n^N \widetilde{X}^n (\rm Ran \mit
F_{\pi^{-1}\varphi+2n_0}\oplus
 \rm Ran \mit F_{\pi^{-1}\varphi+2n_0+2})\}
 \end{eqnarray}
are the bounded operators.

 Put by definition
\begin{equation}
 {\tilde{\mathcal{H}}}_\varphi^{n_0+1}= \overline{\rm Lin \mit
 \{\bigcup \limits_n \widetilde{X}^n (\rm Ran \mit F_{\pi^{-1}\varphi+2n_0}\oplus
 \rm Ran \mit F_{\pi^{-1}\varphi+2n_0+2})\}}.
 \end{equation}
Now the operators \(\tilde{X},\tilde{Y}\) are defined on
 \begin{eqnarray}
 \rm Lin \mit \{\bigcup \limits_{n\geq n_0} \widetilde{\mathcal{H}}_\varphi^{n_0+1}\cap
 F_{\pi^{-1}\varphi+2n}\}.
 \end{eqnarray}
\indent Let us consider the space
\begin{equation}
\Pi_\varphi^{n_0+1}:=\rm Ran \mit
F_{\pi^{-1}\varphi+2n_0+2}\ominus\widetilde{{\mathcal{H}}}_\varphi^{n_0}
\end{equation}
and the space \(\mathcal{H}_\varphi^{n_0+1}:=\rm Lin \mit \{
\bigcup\limits_n \widetilde{X}^n \Pi_\varphi^{n_0+1}\}\) It is
clear that
\begin{equation}
 \tilde{\mathcal{H}}_\varphi^{n_0+1}={\mathcal{H}}_\varphi^{n_0}\oplus
 \mathcal{H}_\varphi^{n_0+1}.
\end{equation}
\indent Continuing this procedure to infinity we will prove the
lemma. \newline

Note that the proof of this lemma is like to the well-known
geometric proof of the theorem about Jordan normal form of
operator [12] .
 \section{Decomposition of the state \(\tau_\varphi\) into the sum of the Gibbs
 states and the end of the proof.}
 Let us decompose the vector \(|\Omega_\varphi\rangle\) into the
 following direct sum
\begin{equation}
|\Omega_\varphi\rangle=\sum \limits_{m=0}^{\infty}|\Omega_{\varphi
m
 }\rangle,
\end{equation}
where \(|\Omega_{\varphi m}\rangle \in
\mathcal{H}_\varphi^{m}.\)\newline It follows from lemma 8 that
for all \( P(X,Y)N_F \in \mathcal{A}^{\star\star}\)
\begin{equation}
\langle\Omega_\varphi|P(\hat{X},\hat{Y})\widehat{N}_F|\Omega_\varphi\rangle=
\sum \limits_{m=0}^{\infty} \langle\Omega_{\varphi
m}|P(\hat{X},\hat{Y})\widehat{N}_F|\Omega_{\varphi m}\rangle.
\end{equation}
Now we state the following\newline
 \indent \textbf{Lemma 9.} The following states
\begin{equation}
\tau_{\varphi n}(P(X,Y)N_F):= \frac{1}{\langle\Omega_{\varphi
n}|\Omega_{\varphi n}\rangle} \langle\Omega_{\varphi n}
|P(\hat{X},\hat{Y})\widehat{N}_F|\Omega_{\varphi n}\rangle
\end{equation}
are well defined and the KMS states.\newline
\indent\textbf{Proof.} Let us show that
\(\tau_n(P(\tilde{X},\tilde{Y})\tilde{N_F})(\varphi)\) are the KMS
states. Let us introduce, the operators
\(\bar{X},\bar{Y},\bar{H}\) defined on
 \(\bigcup \limits_m \{\rm
Ran \mit F_{\pi^{-1}\varphi+2n_0}\oplus...\oplus \rm Ran \mit
F_{\pi^{-1}\varphi+2m+2n_0}\}\) such that

a) The subspaces  \(\mathcal{H}_\varphi^n\) are invariant under
the action of the operators \(\bar{X},\bar{Y},\bar{H}\),

b) The restriction of \(\bar{X},\bar{Y},\bar{H}\) to
\(\mathcal{H}_\varphi^{n_0}\) coincide with the restriction of
\(\tilde{X},\tilde{Y},\tilde{H}\) to \(\mathcal{H}_\varphi^{n_0}\)
respectively, and  the restriction of \(\bar{X},\bar{Y},\bar{H}\)
to\(\mathcal{H}_\varphi^{n_0+m}\) are equal to zero as \(m>0\). We
will find these operators in the following form:
\begin{equation}
\bar{X}=\sum \limits_{m=0}^{\infty}C^1_m\tilde{X}^{m+1}\tilde{Y}^m
F_{\pi^{-1}\varphi+2n_0+2m},
\end{equation}
\begin{equation}
\bar{Y}=\sum
\limits_{m=0}^{\infty}C^2_m\tilde{X}^{m}\tilde{Y}^{m+1}
F_{\pi^{-1}\varphi+2n_0+2m+2},
\end{equation}
\begin{equation}
\bar{N}_F=\sum \limits_{m=0}^{\infty}D_m\tilde{X}^{m}\tilde{Y}^m
F_{\pi^{-1}\varphi+2n_0+2m}.
\end{equation}
It is easy to find such \(C^1_m,C^2_m,D_m\), such that the
restriction of \(\bar{X},\bar{Y},\bar{H}\) to
\(\mathcal{H}_\varphi^{n_0}\) coincide with
\(\tilde{X},\tilde{Y},\tilde{H}\) respectively. It is clear that
these operator are equal to zero on
\(\mathcal{H}_\varphi^{n_0+m}\) \(m>0\). So we have:
\begin{equation}
\tau_0(P(X,Y)N_F)=\langle\Omega_\varphi|P(\bar{X},\bar{Y})\overline{N}_F|\Omega_\varphi\rangle.
\end{equation}
for all \(N_F \in \mathcal{A}^{\star\star}\). Note that the group
of automorphisms \(U_t\) acts on
\(\overline{X},\overline{Y},\overline{N}_F\) as follows
\begin{eqnarray}
U_t(\bar{X})=e^{it}\bar{X},\nonumber\\
U_t(\bar{Y})=e^{-it}\bar{Y},\nonumber\\
U_t(\bar{N}_F)=\bar{N}_F.
\end{eqnarray}
So we have prove the KMS property of the functional \(\tau_0\).
The prove of the KMS property of \(\tau_1,\tau_2,...\) is
analogues to the previous prove.

  Then the following lemma holds.\newline
 \indent \textbf{Lemma 10.} We can chose the set \(A\) such that for all \(a \in
 \mathcal{A}^{\star\star}\)
\begin{equation}
 \tau_{\varphi n}(a)=\rho_{\pi^{-1}\varphi+2n}(a).
\end{equation}
\indent\textbf{Proof.} Let us prove that \(\tau_{\varphi l}(a)\)
\(l=0,2...\) is defined by the Gibbs formula. Note that
 the Hilbert space \(\mathcal{H}_\varphi^{n_0+l}\), is isomorphic to
 \(\Gamma \otimes \bar{V}_{\pi^{-1} \varphi+2n_0+2l}\)
 Here \(\Gamma\) is a some Hilbert space, and \(\otimes\) means
 the tensor product of Hilbert space. The domain of restriction of operators \(\tilde{X},\tilde{Y},\widetilde{F(H)}\)
 to \(\mathcal{H}_\varphi^{n_0+l}\) consider with
 \(\Gamma \otimes {V}_{\pi^{-1}\varphi+2n_0+2l}\). Here  \(\otimes\)
 means an algebraic tensor product and operators \(\tilde{X},\tilde{Y}, \widetilde{F(H)}\)
 at this representation have the form
 \(\mathbf{1}\otimes \hat{X}_\lambda,\mathbf{1}\otimes \hat{Y}_\lambda,\mathbf{1}\otimes (\hat{N}_F)_\lambda
 \), Here \(\hat{X}_\lambda,\hat{Y}_\lambda,(\hat{N}_F)_\lambda\) ---
 are the representations of elements
 \(X,Y,N_F\) in \(V_\lambda\), where \(\lambda=\pi^{-1} \varphi+2n_0+2l\).

 Now \(\forall a=P(X,Y)N_F \in \mathcal{A^{\star\star}}\)
 \begin{eqnarray}
 \tau_l(P(X,Y)N_F)=\sum \limits_{n=0}^{\infty}
 \langle\Omega_{\varphi l}|(P(\tilde{X},\tilde{Y})\widetilde{F(H)}F_{\varphi+2n_0+2l}|\Omega_{\varphi l}\rangle=\nonumber\\
=\sum \limits_{n=0}^{\infty}((P(\hat{X}_\lambda,\hat{Y}_\lambda)
\hat{N}_F)_{n,n} \langle\Omega_{\varphi
0}|F_{\varphi+2n_0+2l}|\Omega_{\varphi 0}\rangle.
\end{eqnarray}
Here symbol \(()_{n,m}\) --- means the matrix element between the
vectors \(|\lambda,n\rangle,|\lambda,m\rangle\). We must prove
that \(\langle\Omega_{\varphi
0}|F_{\varphi+2l+2n})|\Omega_{\varphi 0}\rangle\) is proportional
to the Gibbs weight. It is easy to do by considering the element
\(\langle\Omega_{\varphi
0}|\hat{X}F_{\lambda+2n}\hat{Y}|\Omega_{\varphi 0}\rangle\) and
using the KMS property. So our lemma is proved.
\newline \indent

 \indent So, we see that for all \(a \in
 \mathcal{A}^{\star\star}\)
\begin{eqnarray}
\tau_\varphi(a)=\sum \limits_{i=0}^{\infty} m_i(\varphi)
\rho_{\pi^{-1}\varphi+2i}(a),
\end{eqnarray}
where by definition \(m_i(\varphi)=\langle\Omega_{\varphi
n}|\Omega_{\varphi n}\rangle\). Let us consider the measure
\(d\sigma(\lambda)\) which coincides with \(
d\mu(\pi(\lambda-2k))m_k(\pi(\lambda-2k))\) on each interval \(
[k,2k+2)\). So our state can be represented as
\begin{equation}
\rho(a)=\int \limits_{0}^{+\infty}d\sigma(\lambda)\rho_\lambda(a).
\end{equation}

Now let \(a=N_{e^{itx}} \in A^*\), where \(t \in \mathbf{Q}\). Let
\(\eta_n(x) \in B\) be a sequence of continuous functions such
that \(\rm supp \mit \eta_n(x) \in [-2n,2n]\),
\(\eta_n(x)|_{[-n.n]}=1\), \(0\leq\eta_n(x)\leq1\). We have

\begin{eqnarray}
\int_0^{+\infty}d\sigma(\lambda)\rho_\lambda(N_{\eta_n}N_{e^{itx}})=
\rho(N_{\eta_n}N_{e^{itx}})\label{EQW},
\end{eqnarray}
Where \(\rho_\lambda\) defined in (\ref{32},\ref{33}). The right
hand side of this equality can be represented as follows
\begin{eqnarray}
\rho(N_{\eta_n}N_{e^{itx}})=\int_{-\infty}^{+\infty}d\mu(x)e^{itx}\eta_n(x)
\end{eqnarray}
for some measure \(d\mu\) decreasing faster than any inverse
polynomial. So the right hand side of (\ref{EQW}) tends to
\(\rho(N_{e^{itx}}\) as \(n\rightarrow \infty\). The integrand in
the left hand side of (\ref{EQW}) satisfy
\(|\rho_\lambda(N_{\eta_n}N_{e^{itx}})|\leq 1\) and \(\lim
\limits_{n\rightarrow \infty}
\rho_\lambda(N_{\eta_n}N_{e^{itx}})=\rho_\lambda(N_{e^{itx}})\).
So
\begin{eqnarray}
\int
\limits_{-\infty}^{+\infty}d\sigma(\lambda)\rho_\lambda(N_{e^{itx}})=\rho(N_{e^{itx}})
\end{eqnarray}
or
\begin{eqnarray}
\sigma (\{0\})+\frac{1-e^{-\beta}}{1-e^{-\beta+it}}\int
\limits_{+0}^{+\infty}d\sigma(\lambda)e^{it\lambda}=\int_{-\infty}^{+\infty}d\mu(x)e^{itx}\eta_n(x).
\label{fin}
\end{eqnarray}
Booth sides of equality (\ref{fin}) are continuous on \(t\) so
(\ref{fin}) is valid for arbitrary \(t\). It follows from
(\ref{fin}) that \(d\sigma\) is a linear combination of
\(d\mu(\lambda)-\sigma({0})\delta(\lambda)\) and
\(d\mu(\lambda-2)-\alpha({0})\delta(\lambda-2)\) therefore
\(d\sigma\) decrease faster than any inverse polynomial.

Therefore the part "only if" is proved.
\section{Uniqueness } We can prove the uniqueness of \(\rho\)
 by induction on the number \(B,B^+\) by using the KMS property. Let
 \(\frac{1}{\sqrt{2}}B:=Y\),
 \(-\frac{1}{\sqrt{2}}B^+:=X\).
  consider the following expression
 \begin{equation}
 \rho(B^\pm,...,B^\pm N_F),
 \end{equation}
 where the number of elements \(B^\pm\) is equal to \(n\). The base
 of induction (\(n=0\)) is obvious.
 Suppose that the statement is proved for  \(m=n-1,n-2,...1\).
 Consider the expression
\begin{equation}
 \rho(B^+A N_F),
 \end{equation}
Here \(A\) is a product of \(n-1\) operators \(B^\pm\). Using the
KMS property we have:
\begin{eqnarray}
\rho(B^+A N_F)=e^{-\beta}\rho(B^+A
N_{T_{-2}F})+e^{-\beta}\rho([A,B^+]N_{T_{-2}F}).
\end{eqnarray}
 Iterating this identity we find:
\begin{eqnarray}
\rho(B^+A N_F)=e^{-\beta k}\rho(B^+AN_{T_{-2k}F} )+\sum
\limits_{j=1}^{k}e^{-\beta j}\rho([A,B^+]{T_{-2j}F}).
\end{eqnarray}
We can represent \(\rho(aN_f b)\) \(a,b \in \mathcal{D}\) as an
integral by some measure which decrease faster then any inverse
polynomial. So the first term tends to zero as \(k^Me^{-\beta k}\)
as \(k\rightarrow \infty\) for some integer \(M\). The sum in the
second term has the limit because \(\rho([A,B^+]{T_{-2j}F})\)
tends to zero as \(j^Me^{-\beta j}\) as \(j\rightarrow \infty\)
for some integer \(M\). So
\begin{eqnarray}
\rho(B^+A N_F )=\sum \limits_{j=1}^{\infty}e^{-\beta
j}\rho([A,B^+]{T_{-2j}F}).
\end{eqnarray}
But \([A,B^+]{T_{-2j}F}\) contains only \(n-1\) operators
\(B^{\pm}\). We can write analogues for \(\rho(BA N_F)\). These
representations prove the uniqueness of \(\rho\).
 \section{Conclusion.}
 In the present paper we have investigated the structure of
 Kubo-Martin-Shwinger states on universal enveloping algebra of \(\mathfrak{sl}(2,\mathbf{C})\).
 It is interesting to generalize our results to the infinite
 dimensional case, general Lie algebras and quantum groups.
 \section{Acknowledgements.}
 I wold like to thank I.V. Volovich, L. Accardy, A.N. Pechen and R.A.
 Roschin fo very useful discussions.\newline
 \indent This work was partially supported by the Russian
 Foundation of Basis Reasearch (project 05-01-008884), the grand
 of the president of the Russian Federation (project
 NSh-1542.2003.1) and the program "Modern problems of
 theoretical mathematics" of the mathematical Sciences department
 of the Russian Academy of Sciences.

  \end{document}